# $Co_{25}Fe_{75}$ Thin Films with Ultralow Total Damping

Eric R. J. Edwards, Hans T. Nembach, Justin M. Shaw

Quantum Electromagnetics Division
National Institute of Standards and Technology
325 Broadway
Boulder, CO 80305

## Abstract

We measure the dynamic properties of $Co_{25}Fe_{75}$ thin films grown by dc magnetron sputtering. Using ferromagnetic resonance spectroscopy, we demonstrate an ultralow *total* damping parameter in the out-of-plane configuration of < 0.0013, whereas for the in-plane configuration we find a minimum *total* damping of < 0.0020. In both cases, we observe low inhomogeneous linewidth broadening in macroscopic films. We observe a minimum full-width half-maximum linewidth of 1 mT at 10 GHz resonance frequency for a 12 nm thick film. We characterize the morphology and structure of these films as a function of seed layer combinations and find large variation of the qualitative behavior of the in-plane linewidth vs. resonance frequency. Finally, we use wavevector-dependent Brillouin light scattering spectroscopy to characterize the spin-wave dispersion at wave vectors up to 23 $\mu m^{-1}$.

## Introduction

Emerging applications in magnonics [1], spin-orbitronics [2] as well as fundamental studies in linear and non-linear magnetization dynamics [3] rely on materials that exhibit low values of the magnetic damping parameter. Insulating ferromagnets and ferrites – lacking a significant electron-magnon relaxation channel – have historically exhibited the lowest magnetic dissipation at room temperature. However, many applications require the use of charge currents through the magnetic layer, which excludes the use of insulators. Half-metallic Heusler compounds are predicted to exhibit ultra-low values of the damping parameter as has been observed in a few materials [4]. Such Heusler compounds typically require high temperature processing and/or growth by molecular beam epitaxy. These growth requirements impose a limitation to applications. Recently, it was discovered that $Co_{25}Fe_{75}$ alloys have an ultra-low intrinsic damping parameter of approximately 0.0005, far below previously published values measured in metallic ferromagnets [5] and commensurate with values measured in thin-film ferrites [6, 7]. However, in that work, the total damping was higher than the measured intrinsic damping due to spin-pumping losses into the seed and capping layers [8-10]. For most applications, it is not just the intrinsic damping but the total damping that determines device performance. One means to exclude the spin pumping contributions is epitaxial growth on insulating or semiconducting substrates. Indeed, two recent studies reported the growth of $Co_{25}Fe_{75}$ on MgO and MAO resulting in values of the damping parameters around 0.001 [11, 12]. Removing the constraint of epitaxial growth on single crystalline substrates while maintaining low magnetic damping and high film uniformity may enable integration of this material in a wide range of spintronic and magnonic devices [1,2].

In this work, we demonstrate polycrystalline $Co_{25}Fe_{75}$ thin films grown by dc magnetron sputtering at room temperature on Si/SiO2 substrates with ultralow *total* damping, quantified by a total damping parameter < 0.0013 and inhomogeneous linewidth broadening < 0.5 mT. This is achieved by careful choice of seed and capping layers that provide excellent growth of the CoFe, but at the same time minimize the spin-pumping contribution to the damping. The minimization of inhomogeneous linewidth broadening in these films is also significant since uniformity of element performance determines the overall device figures of merit [13]. Precise control of the material properties in a low damping metallic ferromagnet will enable studies into the fundamental origins of damping due to electron-magnon and magnon-magnon interactions [21,23].

## **Experiment**

Samples were grown by dc magnetron sputtering in a chamber with a base pressure of ≈ 1 × $10^{-7}$ Pa (1 × $10^{-9}$ Torr), wafer to target distance of ≈ 24 cm, wafer to gun inclination angle of 30°, and the sample at ambient temperature during all stages of growth. Deposition rates were calibrated with x-ray reflectometry (XRR) to an accuracy of a few percent. All samples have been grown starting from a thermally grown, 150 nm thick layer of SiO2 on a Si substrate bought from a commercial vendor. The sample holder is rotated at a frequency of 0.5 Hz during deposition to minimize in-plane variation of magnetic parameters across the wafer. All materials except for select Cu layers were sputter deposited in pure Ar atmosphere at a pressure of ≈ 60 mPa (0.5 mTorr). In the case of Cu, we have sputter deposited some layers with ≈ 5 % g/g $N_2$ in the gas mixture, which we denote in the following as Cu(N) to distinguish it from those Cu layers grown in pure Ar. All samples have been capped by a 5 nm thick layer of Al to protect against ambient conditions. The crystal structure was characterized by in-plane x-ray diffraction (XRD) using parallel beam optics with a Cu Kα radiation source, and the film roughness has been characterized by atomic-force microscopy (AFM).

We use broadband, field-swept vector network analyzer ferromagnetic resonance (VNA-FMR) in the out-of-plane configuration with the external magnetic field oriented perpendicular to the film plane to measure the total Gilbert damping α and inhomogeneous linewidth broadening $\Delta H_0$ of these films by fits of the frequency-dependent resonance linewidth to the equation:

$$\Delta H = \Delta H_0 + \frac{4\pi \alpha f}{\gamma \mu_0}.$$

Exemplary resonance spectra are displayed in Figure 1. For details of the fitting procedure concerning also the determination of spectroscopic g-factor *g* and effective magnetization $M_{eff}$ we refer to [14]. For these measurements, we utilize a room-temperature bore superconducting magnet capable of generating magnetic fields of 3 T in a 1 $cm^3$ volume at the sample position. To minimize the contribution of the spatial variation of the external field to the inhomogeneous linewidth broadening, we cleave all samples post-deposition to dimensions of approximately 5 mm X 5 mm. To minimize the extrinsic radiative damping [15], we perform all measurements on films of thickness larger than 3 nm with a 70 μm thick spacer of sapphire inserted between the sample and the waveguide [5]. All samples have been spin-coated with a thin layer of PMMA prior to measurement. Both the out-of-plane and in-plane measurements have been performed with a coplanar waveguide (CPW) with a center conductor width of 150 μm and 50 Ω characteristic impedance. Power dependent measurements have been performed to verify linear response behavior of the magnetic susceptibility and we apply no more than 0.3 mW (-5 dBm) at the output port of the VNA during measurement. The magnetic field is recorded during field sweeps by a calibrated Hall probe.

Propagating spin-wave modes were detected by use of a wavevector-dependent Brillouin light scattering spectrometer (BLS). For details of the apparatus, we refer the reader to [16, 17]. The sample is mounted on a rotary stage to set the angle of incidence of the probe light. By scanning the angle of incidence from 9° to 81°, we scan the measured spin-wave wavevector in the range 3.69 µm$^{-1}$ to 23.33 µm$^{-1}$. The external field is oriented perpendicular to the plane of incidence of the probe light with a magnitude of 0.13 T. This geometry results in measurement of the so-called Damon-Eshbach modes. The recorded spectra are fit according to the procedure described in [18], which allows extraction of the resonance frequency of the mode as a function of spin-wave wavevector.

## Results

In the work by Schoen et al.[5], high quality $Co_{25}Fe_{75}$ with an exceptionally low intrinsic damping parameter and low inhomogeneous linewidth was achieved using a Ta/Cu seed layer and a symmetric Cu/Ta capping layer. However, the use of Ta in the seed layer induces a large spin-pumping contribution to the damping [19]. As a result, even though the intrinsic damping was reported to be exceptionally low, the total damping parameter was increased by use of the Ta/Cu seed/capping layer. To obtain magnetic damping near the intrinsic limit set by the material, it is necessary to lower both the spin current absorption in the seed/capping layers and maintain low spin-memory loss at the interfaces with the ferromagnet. In addition to these requirements there are two other considerations that need to be taken into account: (1) The seed layer must have good adhesion properties while simultaneously promoting good growth of the materials deposited on top of it. Ta is exceptional at this task, motivating its use in the initial study by Schoen et al.; (2) the seed layer must provide good growth of the magnetic layer. We have found from our previous experience, that growth of magnetic 3d transition metals on Cu templates reliably results in high quality materials with good magnetodynamic properties [26,27].

With these considerations in mind, we modify the seed layer by using 3Ti/3Cu since Ti layers of this thickness are not expected to efficiently absorb spin current, while keeping the Cu layer to provide a good template for the CoFe growth. Since the primary purpose of the capping layer is to prevent oxidation of the CoFe and adhesion and growth are not an issue for the capping layer, other materials can be considered. A quick survey of materials (not shown) indicated that a 5 nm Al capping layer minimized the spin-pumping contributions. The self-passivation of a thin (2 nm to 3 nm) AlOx layer formed when the Al layer is exposed to ambient conditions provides long-term protection of the rest of the stack. Finally, we also investigate the use of Cu(N) instead of Cu since adding a few percent of $N_2$ in the gas mixture used for sputtering results in smoother Cu layers.

Figure 2 is a plot of the measured total damping parameter in the out-of-plane geometry as a function of 1/t where t is the thickness of the CoFe layer. Also included in the plot is the original data from Schoen et al. where a symmetric Ta/Cu seed/capping layer was used. First, we note that thickness dependence of the damping is greatly reduced in the seed layers that use Ti versus Ta. This indicates that the spin-pumping contribution is dramatically reduced with the use of the Ti/Cu or Ti/Cu(N) seed and Al capping laibyers. In fact, the reduction of the spin-pumping losses yields a value of 0.0018 for the damping parameter in a 2 nm CoFe layer, which is exceptional for such a thin layer. For comparison, a 2 nm CoFe layer has a damping parameter of 0.0054 with the use of a Ta/Cu seed and cap. Our results are consistent with a recent report on ultrathin CoFeB layers that shows that damping can be minimized by careful engineering of the spin-pumping contribution from Ta [20]. Figure 2 also shows a minimum damping parameter of 0.0012 for the 10 nm layer with the Ti/Cu(N). Of critical importance to applications is also the exceptionally low value of 0.4 mT for the inhomogeneous contribution to the linewidth. This yields a total linewidth of 1.0 mT at 10 GHz, which is a typical figure of merit.

Figure 2 also shows that there is no significant difference between the total damping parameters obtained in $Co_{25}Fe_{75}$ grown on Ti/Cu or Ti/Cu(N).  Although there are no discernable differences in relaxation properties in the out-of-plane geometry, the situation is different when the magnetic field is applied in the plane of the film.  Figures 3a and 3b show the linewidth versus frequency in both the in-plane and out-of-plane measurements for the Ti/Cu seed and Ti/Cu(N) seed samples, respectively. In the case of the Ti/Cu(N) seed layer, there is a small increase in the damping parameter to a value of 0.0016 in the in-plane geometry and negligible increase in inhomogeneous linewidth broadening. In the case of the Ti/Cu layer, there is a more significant increase in the linewidth. Both the slope (damping) and the y-intercept (inhomogeneous linewidth broadening) have increased.  Upon closer inspection, the in-plane linewidth data is not linear over the entire frequency range and has an inflection point at approximately 28 GHz. While it is subtle in the 10 nm sample, this behavior is more pronounced in thicker samples (not shown).  Such behavior is characteristic of 2-magnon scattering [21] and is consistent with increased roughness in the Ti/Cu seed layer.  This is an important consideration if an application requires a material that is in-plane magnetized.

Finally, to verify the expected characteristics of propagating spin wave modes important for magnonics applications, we have measured the resonance frequency as a function of spin-wave wavevector using BLS as described above. From the measured dispersion relation, we extract the group velocity. Our results are plotted in Figure 4. The observed dispersion relation is consistent with the theory of Damon-Eshbach waves in this wavevector regime, where the relative quadratic contribution of the exchange interaction is small [22]. At the external field magnitude of 0.13 T, we obtain a maximum spin-wave group velocity of 5250 km/s, monotonically decreasing with increasing wavevector up to the experimental limit.

## Discussion

Our results demonstrate the importance of considering not just the intrinsic damping of the magnetic material, but also the spin-pumping effects into the adjacent materials, which can dominate the damping in some cases.  Especially in the case of thin layers important to many spintronic devices, this extrinsic loss mechanism will dominate the total magnetic damping. While we have shown a particular seed layer and capping layer combination that yields ultra-low total damping in thin films down to 2 nm thickness, these seed/capping layers may not be compatible with a particular application or process. The question remains if other seed layers can also be used to produce similar results, if a seed layer is needed at all, and what effect the seed layer has on the intrinsic damping.  While a full, comprehensive study of all possible seed layers is impossible for this study, we do investigate several alternatives which include: no seed layer, Cu, Ti/Al, Ta, Ta/Cu and Ti.  With the exception of Ta and Ta/Cu, all of these possibilities should have minimal spin-pumping contributions.  Figure 5a shows a comparison of the measured damping for each of these seed layers for a 10 nm thick $Co_{25}Fe_{75}$ layer. Values of the damping parameter vary from approximately 0.0010 to 0.0014 except for the samples that contain Ta in the structure, which are higher. These results further demonstrate the importance of the spin-pumping contributions to the damping. It is interesting to note that the samples with the two lowest values of the damping parameter are Ti and Ti/Al seed layers, however the picture becomes more complicated when the inhomogeneous contribution is taken in consideration as shown in Figure 5b.

While the damping parameters are relatively similar with the different seed layers, there is a dramatic variation of $\Delta H_0$ with different seed layers. Large differences with use of different seed layers are further observed in the in-plane FMR spectra. These are summarized in Figure 6 that show the linewidth versus frequency plots from FMR data taken in both the in-plane and out-of-plane geometries. Most seed layers show considerable increases in the linewidth in the in-plane geometry relative to the out-of-plane geometry. In addition, the non-linear linewidth versus frequency that is characteristic of 2-magnon scattering is also prevalent in most in-plane measurements [21].

We performed XRD analysis of the $Co_{25}Fe_{75}$ to determine the lattice constant and crystalline texture that results from the use of various seed layers. The use of Ta/Cu, Ti/Cu, and Ta seed layers provides good [011] texture to the $Co_{25}Fe_{75}$. However, no correlations were found between the quality of the Co25Fe75 (as evaluated from the relaxation properties and inhomogeneous linewidth) and either the lattice constant or the crystalline texture.

An interesting observation is that the samples with the lowest values of damping (Ti and Ti/Al) also have the highest values of $\Delta H_0$. Such an inverse relation between alpha and $\Delta H_0$ might be an artifact that results from fitting the linewidth data over too short of range in frequency or over a range of lower frequencies. For example, if there are low-field loss effects or 2-magnon scattering effects, over a small range in frequency the data can appear linear, but yield an erroneous value of $\alpha$ and $\Delta H_0$. However, in the present case, there is a physical reason why there may be an inverse relation between $\alpha$ and $\Delta H_0$. Figure 7 shows a schematic diagram (adopted from Ref. [23]) showing a plot of the intrinsic damping parameter as a function of $1/\tau$ where $\tau$ is the electron-lattice scattering time. Such a curve is derived from theory that includes both interband and intraband scattering mechanism and we refer the reader to references since a detailed discussion is beyond the scope of this work [23-24]. It is important to point out the two regimes where intraband scattering dominates (the left side of the plot where $\tau$ is large) and the regime where interband scattering dominates (the right side of the plot where $\tau$ is small). The scattering time is related to both scattering from thermal fluctuations (temperature) as well as structural disorder. So, in the case of increased structural disorder, the value of $\tau$ will decrease (i.e., scattering becomes more frequent). From the previous work on $Co_{25}Fe_{75}$, it was determined that the damping is dominated by intraband scattering. As a result, an increase in structural disorder can lead to a decrease in the intrinsic damping of the material.

We therefore explore this possibility by systematically varying the structural disorder. A series of 2 nm thick $Co_{25}Fe_{75}$ layers were deposited on a 3Ti/(x)Cu seed layer where x was varied from 2 nm to 30 nm. The different thicknesses of Cu change the roughness of the surface for which the magnetic layer is deposited. As previously reported, this variation in roughness can affect the magnetic homogeneity of the resulting magnetic layer [13,25]. The roughness of this series of samples was measured by AFM and is plotted in Figure 8a as a function of the Cu thickness. While there is considerable scatter in the data, the expected trend in the data is clear: a thicker Cu layer produces a rougher surface. The values of $\alpha$ and $\Delta H_0$ are plotted in Figure 8b. The damping parameter monotonically decreases from 0.0019 to 0.0012 as the thickness of the Cu layer is increased from 2nm to 15 nm. In contrast, the value of $\Delta H_0$ monotonically increases from about 1 mT to 9 mT over the same range. (Values for of $\alpha$ and $\Delta H_0$ for the 30 nm sample are not reported since the exceptionally large $\Delta H_0$ did not produce reliable fits to the data).

As previously discussed, this inverse relationship between α and $\Delta H_0$ merits suspicion and therefore requires careful consideration of the raw data. The linewidth versus frequency data that was used to determine α and $\Delta H_0$ are plotted in Figure 8c. These data show excellent linearity over a large range in frequency making the results in Figure 8b rather convincing. In fact, this may provide insight into the data in Figure 2 that shows in increase in the y-intercept for the Ti/Cu and Ti/Cu(N) seed layers compared to Ta/Cu. These data indicate that the intrinsic damping for CoFe grown on Ti/Cu is closer to 0.0010 than 0.0005 ± 0.0003 as previously determined for Ta/Cu, which is also included in the plot [5]. More evidence of this is given in [11] where it is shown that the better lattice matched epitaxially grown CoFe has a higher damping parameter than similar films grown on substrates with poorer lattice matching, increasing the number of dislocation defects. These results suggest that variation of the intrinsic damping of $Co_{25}Fe_{75}$ may also contribute to the variation of the total damping. A systematic investigation needed to further support this hypothesis is left for future work.

## **Conclusion**

In conclusion, we have demonstrated that a total damping of 0.0013 is obtainable in films grown by standard sputtering deposition techniques by minimizing spin pumping contributions to the damping. To the best of our knowledge, these values are the lowest reported to date in this system. At the same time, we have demonstrated that ultralow total linewidth films can be grown with these techniques. The simultaneous optimization of damping and inhomogeneous line broadening will enable the use of this material in applications and fundamental studies of damping in metallic ferromagnets. Our work indicates that a detailed understanding of the intrinsic damping variation as a function of growth conditions may be a route to further reduction of the damping of metallic ferromagnets. Finally, this work emphasizes the importance of the seed layer and growth conditions for the quality of the $Co_{25}Fe_{75}$ film.

# Figures

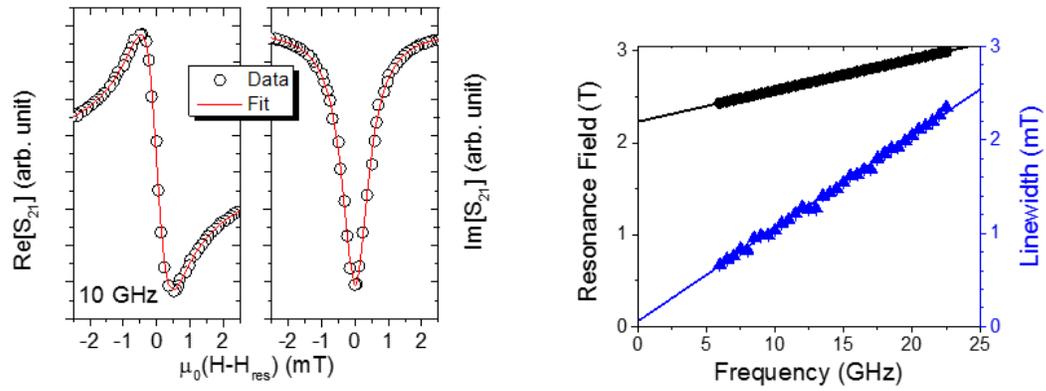

**Fig. 1**. (a) Exemplative real and imaginary ferromagnetic resonance data measured with VNA-FMR on a 3Ti/5Cu/10Co$_{25}$Fe$_{75}$/5Al sample. (b) Extracted resonance field and linewidth as a function of frequency for the same sample.

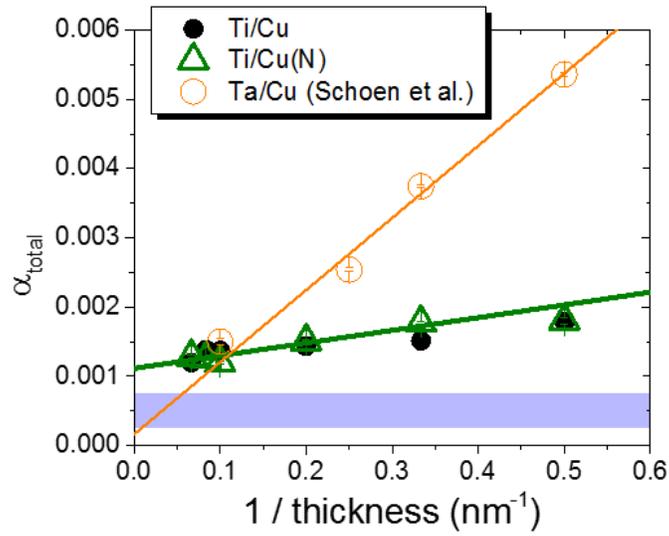

**Fig. 2**. Ferromagnetic resonance linewidth versus frequency for 10 nm of $Co_{25}Fe_{75}$ on a selection of seed layers. The reduced slope of the Ti/Cu and Ti/Cu(N) samples relative to Ta/Cu indicates a lower contribution to the total damping from the spin pumping mechanism. The blue filled region indicates the intrinsic damping measured in [5].

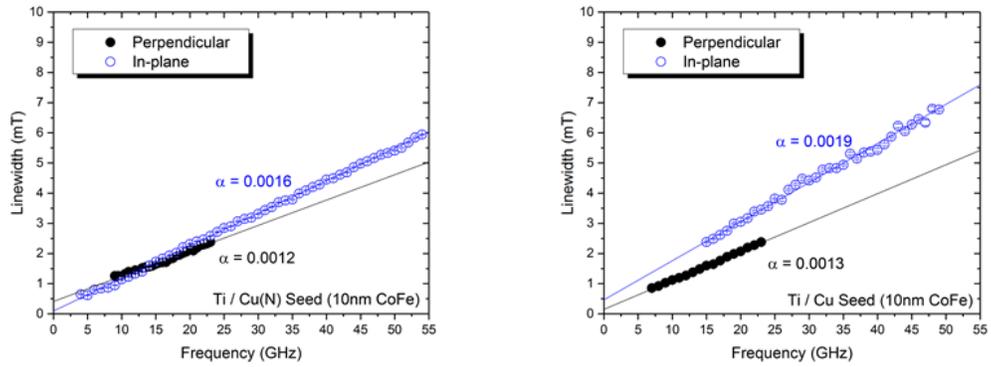

**Fig. 3**. Ferromagnetic resonance linewidth data in the in-plane and out-of-plane measurement configurations for Ti/Cu(N) and Ti/Cu seed layers at 10 nm of $Co_{25}Fe_{75}$.

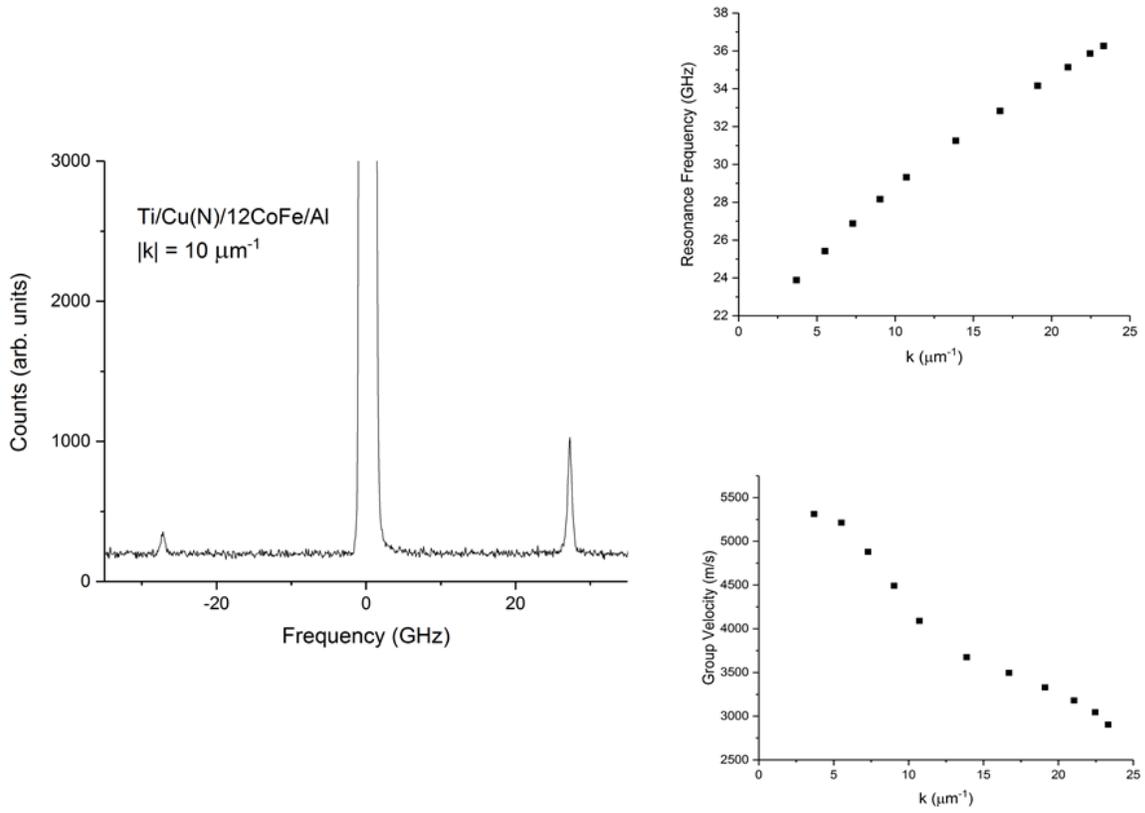

**Fig. 4**. (a) Exemplary Brillouin light scattering spectrum for a Ti/Cu(N) seed layer at 12 nm of $Co_{25}Fe_{75}$. (b) Extracted resonance frequency versus spin-wave wavevector for the same sample for surface waves. (c) The spin-wave group velocity determined from the spin-wave dispersion.

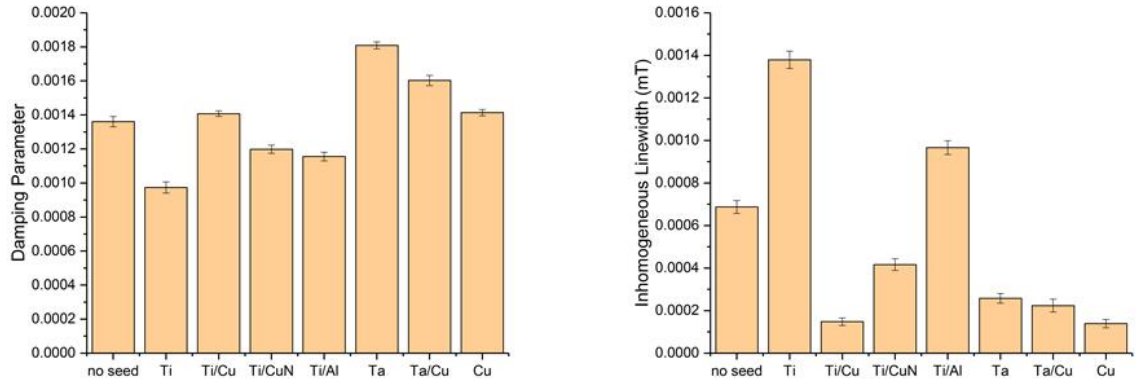

**Fig. 5**. (a) The total damping parameter determined from out-of-plane ferromagnetic resonance measurements for a selection of seed layers for 10 nm of $Co_{25}Fe_{75}$. (b) The inhomogeneous linewidth broadening parameter determined from the same measurements.

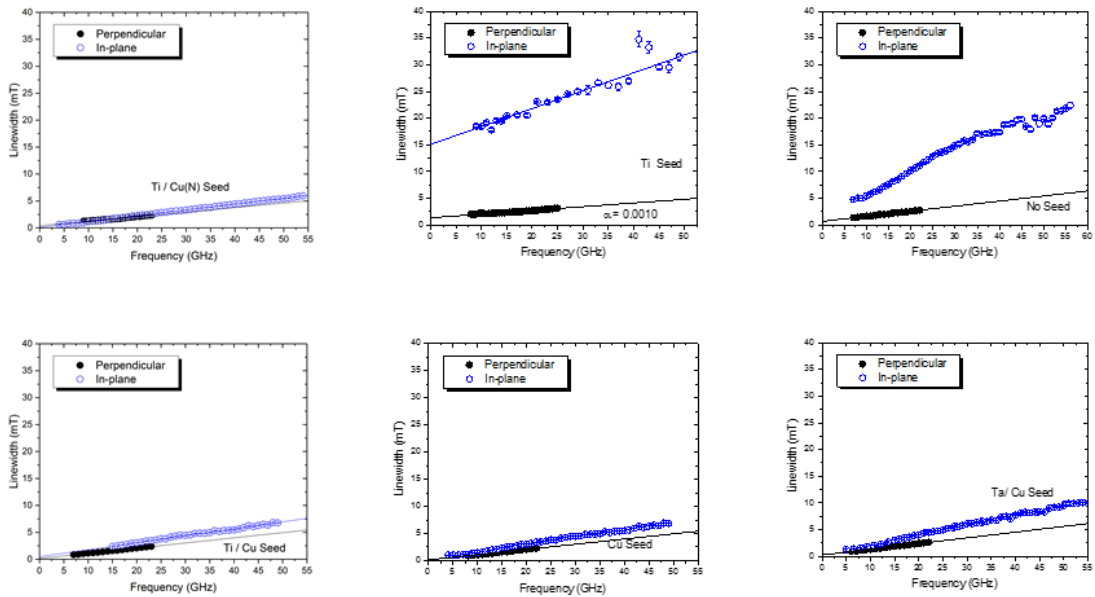

**Fig. 6**. Ferromagnetic resonance linewidth as a function of resonance frequency for a variety of different seed layers at 10 nm of $Co_{25}Fe_{75}$. The large variation in the in-plane linewidth indicates substantial contributions from two magnon scattering for certain seed layers.

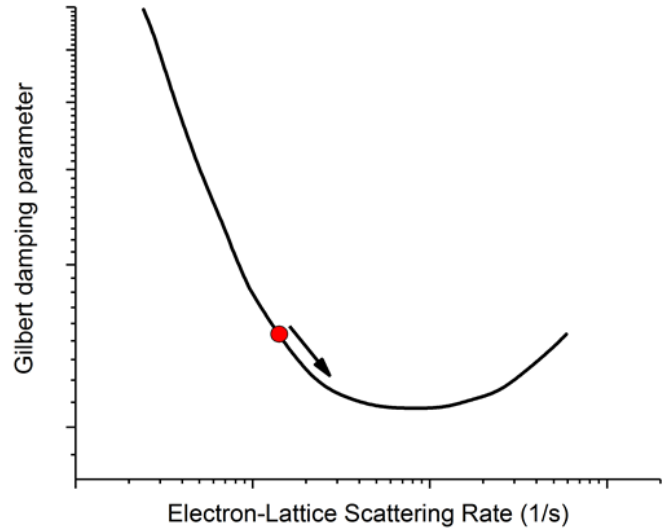

**Fig. 7**. The intrinsic damping as a function of electron-lattice scattering rate, adapted from [23]. The red dot indicates a point located in the so-called intraband regime where an increase in electron-lattice scattering rate can lead to a decrease in the intrinsic magnetic damping.

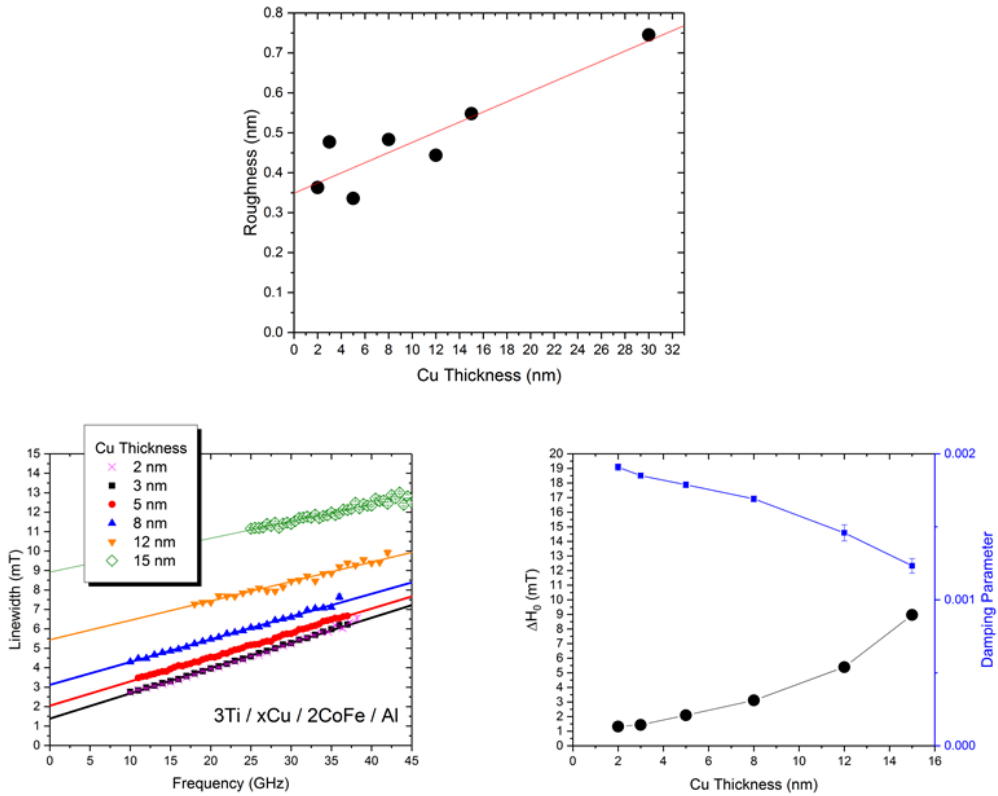

**Fig. 8**. (a) Root-mean-square film roughness determined from atomic force microscopy for a 3Ti/xCu/2Co$_{25}$Fe$_{75}$/5Al stack where x is the varied Cu layer thickness. (b) Ferromagnetic resonance linewidth measured in the out-of-plane configuration as Cu buffer layer thickness is varied. (c) Damping parameter and inhomogeneous linewidth broadening parameter determined from the data of (b).